\documentclass[aps,prl,twocolumn,superscriptaddress,showpacs,longbibliography]{revtex4-1}
\usepackage{bm} 
\usepackage{graphicx}
\usepackage{epsfig}
\usepackage{hyperref}
\usepackage{natbib}
\usepackage{amsmath}
\usepackage{amssymb}
\usepackage{color}

\renewcommand{\i}{\mathrm{i}}

\newcommand{\rmi}{\mathrm{i}}
\newcommand{\e}{\mathrm{e}}

\renewcommand{\Im}{{\rm Im\,}}

\begin{document}

\title{Radiative topological states in resonant photonic crystals}

\author{A.\,V.\,Poshakinskiy}\email{poshakinskiy@mail.ioffe.ru}
\author{A.\,N.\,Poddubny}
\affiliation{Ioffe Physical-Technical Institute of the Russian Academy of Sciences, 194021
St.~Petersburg, Russia}
\author{L.~Pilozzi}
\affiliation{Istituto dei Sistemi Complessi, CNR, C. P. 10,
Monterotondo Stazione, Rome I-00016, Italy}
\author{E.\,L.\,Ivchenko}
\affiliation{Ioffe Physical-Technical Institute of the Russian Academy of Sciences, 194021
St.~Petersburg, Russia}

\begin{abstract}
We present a theory of topological edge states  in one-dimensional resonant photonic crystals with a compound unit cell. Contrary to the traditional electronic topological states the states under consideration are radiative, i.e., they decay in time due to the light escape through the structure boundaries. We demonstrate that the states survive despite their radiative decay and can be detected both in time- and frequency-dependent light reflection.
\end{abstract}








\maketitle

{\it Introduction.}
Topological insulator is an electronic material that has a band gap in its interior like an ordinary insulator but
possesses conducting states on its edge or surface. The surface states of topological insulators have been extensively studied both in 
two- and three-dimensional materials~\cite{RevModPhys.82.3045}.
 Recently an untrivial link has been revealed between such seemingly distinct systems  as topological insulators, {\it one-dimensional} (1D) quasicrystals, and periodic 1D crystals with compound unit cell~\cite{Lang2012,kraus2012,dassarma2013}. Particularly, it has been demonstrated that the 1D Aubry-Andr\'e-Harper (AAH) model, or a ``bichromatic'' system (both incommensurate and commensurate), exhibits topological properties similar to those attributed
to systems of a higher dimension~\cite{Lang2012,kraus2012,dassarma2013}. This model allows states at sharp boundaries between two distinct topological systems. The system is described by a 1D tight-binding Hamiltonian with nearest-neighbor hopping and an on-site potential~\cite{Kraus2012b}. In the generalized AAH model both the hopping terms and the on-site potential are cosine modulated. It is the modulation phase that adds the second degree of freedom and permits one to relate the descendent 1D model with a 2D
``ancestor'' system which has a 2D band structure and quantized Chern numbers. 
In this Letter,  instead of quasiparticles which tunnel from one site to another, we consider a 1D sequence of sites with resonant excitations long-range coupled 
through an electromagnetic field~\cite{Ivchenko2013}. 
Such system is open, its eigenfrequencies are complex and its eigenstates are quasistationary due to the radiative decay.  Hence, the resonant optical lattice stands out of the standard classification of topological insulators, developed for conservative and Hermitian electronic problems~\cite{Shinsei2010}. Nevertheless, we show here that this 1D bichromatic resonant photonic crystal demonstrates the topological properties in spite of being open and formulate general condition for the edge state existence. We also demonstrate how the radiative character of the problem opens new pathways to optical detection of the edge states.
This provides an important insight into the rapidly expanding field of the electromagnetic topological states in photonic crystals~\cite{Haldane2008,wang2008}, coupled cavities~\cite{hafezi2011}, waveguide arrays~\cite{kitagawa2012,rechtsman2013b,rechtsman2013},  and metamaterials~\cite{Khanikaev2013}.

{\it Model.} We consider a 1D resonant photonic crystal consisting of alternating layers $\mathcal A$ and $\mathcal B$. The dielectric constant $\varepsilon_b$ of the material $\mathcal B$ is frequency-independent while the thin layer $\mathcal A$ is characterized by single-pole amplitude coefficients of light reflection and transmission,
\begin{equation} \label{rt}
r_{\mathcal A} (\omega) = - \frac{\rmi \Gamma_0}{\omega - \omega_0 + \rmi (\Gamma_0+\Gamma)} \:, \; t_{\mathcal A}(\omega) = 1 + r_{\mathcal A} (\omega) \:.
\end{equation}
Here, $\omega$ is the light frequency, the resonance frequency $\omega_0$, radiative ($\Gamma_0$) and non-radiative ($\Gamma$) decay rates are three basic parameters of the excitation in a single layer $\mathcal A$ sandwiched between semi-infinite layers $\mathcal B$. This model can be applied to excitonic~\cite{Ivchenko2005}, dielectric and plasmonic multilayers~\cite{taubert2012,Weiss2011}, to coupled waveguides~\cite{yanik2004}, and even to nuclear excitations in multilayers containing different isotopes of the same element, see the review [\onlinecite{Ivchenko2013}]. The multilayer system can be equivalently described by a set of coupled equations for the resonant dielectric polarizations $P_n$ of the
 layer $\mathcal A$ ($n=1,2,...$), as follows
\begin{equation} \label{qw}
( \omega_0 - \omega ) P_n - {\rm i} \Gamma_0 \sum\limits_{n'} {\rm e}^{{\rm i} q |z_{n} - z_{n'}| }  P_{n'} =  0\:,
\end{equation}
where $q = \omega \sqrt{\varepsilon_b}/c$ is the light wave vector in the material $\mathcal B$ and $z_n$ is the center of the $n$-th layer $\mathcal A$~\cite{Ivchenko2005}.
Following Ref.~[\onlinecite{Kraus2012b}] we take a bichromatic structure with the $\mathcal A$ layers centered at
\begin{equation}\label{eq:pos}
z_n = d [ n + \eta \cos \left( 2 \pi b n + \phi \right) ]\:,
\end{equation}
where $b$ is a dimensionless parameter of the system, $d$ is the period in the primary lattice, and $\eta$ is a small modulation amplitude.
Figure 1 illustrates the structure with $b=1/3$ representing a periodic photonic crystal with the period $D=3d$.

\begin{figure}[t!]
  \includegraphics[width=0.95\columnwidth]{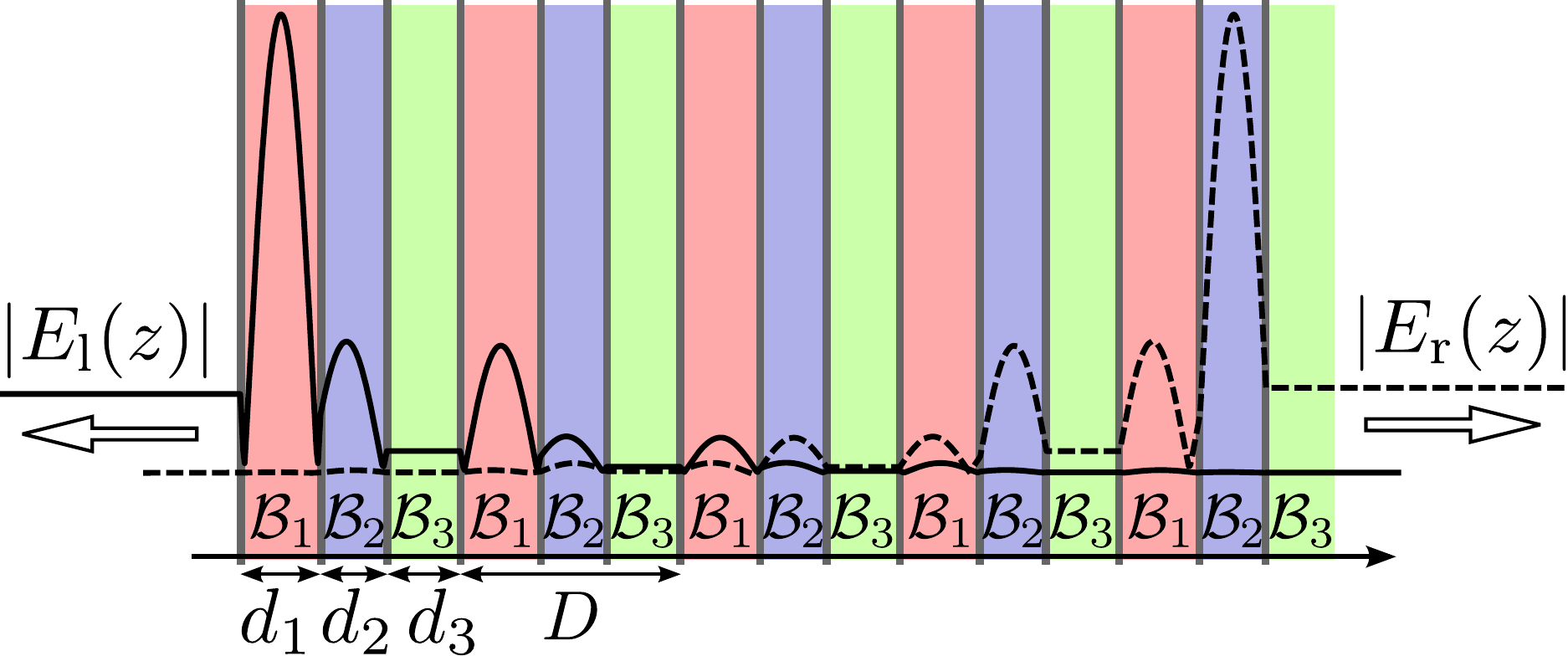}
\caption{(Color online) Illustration of the periodic structure with three layers $\mathcal A$ in the unit cell. Vertical lines indicate the $\mathcal A$ layers, the labels ${\mathcal B}_1$, ${\mathcal B}_2$ and ${\mathcal B}_3$ mark the barriers of different thicknesses $d_l = z_{l+1} - z_l$ where $l=1,2,3$ and $z_4 = z_1 + D$. The solid line shows the electric field distribution for the state localized on the left edge, the dashed line corresponds to the right-edge state. The parameters used are $d = \lambda_0/2$ ($\lambda_0 = 2 \pi c/\sqrt{\varepsilon_b}\omega_0$), $\varkappa=\phi-\pi/6=\pi/2$ and  $\eta=0.2/\pi$.}
\label{fig:states}
\end{figure}

{\it Topological properties of the lattice.} Let us relate the 1D multilayer system with a 2D ``ancestor'' lattice with the sites at $z = z_n$ and $x=m$ $(m=0,\pm1\dots)$, where $x$ is an extra axis. To this end we introduce the site polarization $P_{nm}$ of the 2D lattice replacing $P_n$ by $P_{n,\phi}$ in Eq.~(\ref{qw}), considering $\phi$ as the wave vector component along the $x$ direction and defining $P_{nm}$ by the Fourier transform $P_{n,\phi} = \sum_{m'} {\rm e}^{- {\rm i} m' \phi} P_{n,m'}$. After multiplying  Eq.~(\ref{qw}) by ${\rm e}^{{\rm i} m \phi}/2\pi$ and integrating over $\phi$ from 0 to $2 \pi$ we obtain
\begin{equation} \label{qw3}
\left( \omega_0 - \omega \right) P_{nm} - {\rm i} \Gamma_0\,\sum\limits_{n' m'} \Lambda_{nm;n'm'} P_{n'm'} = 0\:. 
\end{equation}
Here, the coupling coefficients are given by
 \begin{multline} 
\Lambda_{nm;n'm'} = {\rm e}^{{\rm i} q d|n - n'| }\,  {\rm e}^{{\rm i} \pi b (m' - m) (n + n') }\\
  \times \,
  J_{m-m'}[2 \eta q d \sin(\pi b |n - n'|)]\:,\label{Lambda}
\end{multline}
where $J_l$ is the Bessel function of the order $l$. 
These coefficients retain the long-range couplings from Eq.~\eqref{eq:pos}, which is distinct
from the nearest-neighbor AAH model and its  2D counterpart~\cite{Kraus2012b}. In the following we set $b$ in Eq.~(\ref{eq:pos})
to be a rational number ${\mathcal M}/{\mathcal N}$ in which case the structure is periodic with the period $D = {\mathcal N}d$ and contains ${\mathcal N}$ layers $\mathcal A$ in the unit cell.

The presence of edge states is a generic topological property of various AAH models studied so far~\cite{Lang2012,Kraus2012b,dassarma2013}. This is revealed by the nontrivial Chern numbers 
of the allowed zones of the infinite structure. The propagating solutions satisfy the Bloch condition $P_{l + s {\mathcal N}}(k,\phi) = {\rm e}^{{\rm i} skD}P_l(k,\phi)$ where the index $l=1,2...{\mathcal N}$ enumerates the layers $\mathcal A$ in the unit cell, $s = 0,\pm1\dots$ and $k$ is the wave vector $z$-component defined in the interval $( - \pi/D, \pi/D]$. The polarizations $P_l(k,\phi)$ satisfy the equations $\sum_{l'} H_{ll'} P_{l'} = \hbar \Omega P_l$, where the Hermitian matrix
\begin{eqnarray}\label{eq:system2}
 H_{ll'}&=& \hbar \omega_0 \delta_{ll'} \\  &+& \hbar \Gamma_0 \frac{\e^{-\i kD\, {\rm sign}\{l-l'\}} \sin qz_{ll'} + \sin q(D-z_{ll'})}{\cos kD - \cos qD} \:,\nonumber
 \end{eqnarray}
plays the role of the Hamiltonian ($z_{ll'} = |z_{l} - z_{l'}|$). Due to the time-inversion symmetry, the eigenfrequency $\Omega(k,\phi)$ is an even function of $k$. It is convenient to make a shift of the phase in Eq.~(\ref{eq:pos}) replacing $\phi$ by $\varkappa - b\pi+\pi/2$ and defining the ``wave vector'' $\varkappa$ in the interval $(-\pi, \pi]$. The shift allows us to disclose an important symmetry property of the system: the structure corresponding to a particular value of $\varkappa$ is spatially inverted under the reversal $\varkappa \to - \varkappa$. This means that the eigenfrequency $\Omega(k,\varkappa)$ is also even in $\varkappa$. Another property follows from the invariance of the infinite system  under the shift of the numeration $n \to n+p$ in Eq.~\eqref{eq:pos}. This yields the symmetry $\Omega(k,\varkappa+2\pi b p) = \Omega(k,\varkappa)$.

For the Chern number $C_{\nu}$ of the band $\nu$, we use the standard definition
 \begin{equation}\label{eq:chn}
  C_{\nu} =   \int\limits_{-\pi}^{\pi}\frac{d\varkappa }{2\pi\i}   \int\limits_{-\pi/D}^{\pi/D}  dk 
  \left( \partial_k \left< P \bigl| \partial_{\varkappa}  P \right> 
  - \partial_{\varkappa} \left< P \bigl| \partial_k  P \right> \right)\:,
 \end{equation} 
where $P_l(k,\varkappa)$ is the eigensolution in the band $\nu$, $\left< P \bigl| \partial_{\varkappa}  P \right> \equiv \sum_l P_l^*(k,\varkappa) \partial P_l(k,\varkappa) / \partial \varkappa$, and $\left< P \bigl| \partial_k  P \right>$ is defined similarly.
\begin{figure}[t!]
\includegraphics[width=0.95\columnwidth]{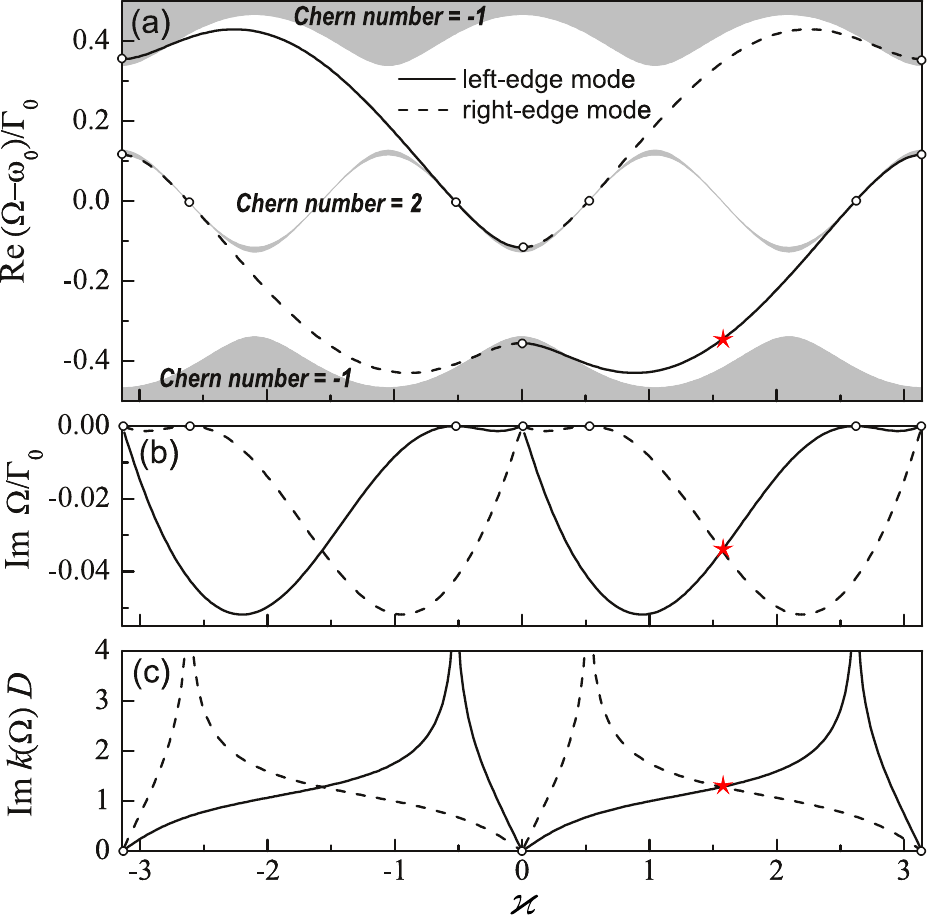}
\caption{(Color online) (a) The band structure as function of the  ``ancestor'' lattice wave vector $\varkappa=\phi-\pi/6$  characterizing the distribution of three $\mathcal A$ layers in the unit cell. The gray regions are the allowed polariton  zones, while the white regions are the stop-bands. The lines show the dependence of the real part of the frequency of the mode localized on the left (solid) and right (dashed) edges of the system. Circles indicate points where an edge modes is absent. The star corresponds to the value of $\varkappa=\pi/2$ chosen for the calculation presented in Fig.~\ref{fig:refl}. (b) Dependence of the edge mode decay rate $\Im\Omega$ on the parameter $\varkappa$. (c) The $\varkappa$-dependence of the eigenstate spatial decay constant $\Im \{k(\Omega)\}D$. The calculation is performed for $d = \lambda_0/2$, $b = 1/3$, $\eta=0.2/\pi$, and in the absence of non-radiative damping.
}
\label{fig:zones}
\end{figure}

As shown below, the structure must lack an inversion center in order to have nontrivial Chern numbers and topological edge states. This requires at least 3 layers $\mathcal A$ in the unit cell. Fig.~\ref{fig:zones}(a) presents the dependence of the edges of allowed zones (gray areas) on the wave vector $\varkappa$ for the lattice with $b=1/3$,
${\mathcal N} = 3$, $D=3d$ and the primary period satisfying the resonant Bragg condition $d = \lambda_0/2$~\cite{Ivchenko2013}. The corresponding Chern numbers are equal to $-1$, $2$ and $-1$. Real parts of the eigenfrequencies of two edge states of the structure are depicted by solid and dashed lines in Fig.~\ref{fig:zones}(a). Next we give the details of how these states are found and propose methods of their detection.

{\it Radiative edge states.}
Direct calculation of the  eigenfrequencies $\omega$ from Eq.~\eqref{qw}  is a numerically challenging problem of solution of a transcendent equation (\ref{qw}) where $\omega$ enters the phase factors $\e^{\i qd|z_n-z_{n'}|}$ via $q=\omega \sqrt{\varepsilon_b}/c$. Instead, we study the properties of the structure reflection coefficient $r(\omega)$ as a function analytically continued onto the complex plane $\omega = \omega' + {\rm i} \omega''$. As an additional advantage, the coefficient $r(\omega)$ is directly accessible in experiments on photonic crystals and can be readily evaluated using the transfer matrix technique~\cite{Ivchenko2005}. It is instructive to start from the analytical properties of the reflection coefficient $r_{\infty}(\omega)$ from the semi-infinite structure. This function of $\omega$ has poles indicating the edge states and discontinuities across the branch cuts on the real axis related to the allowed bands of the corresponding infinite structure. In the reflection coefficient $r(\omega)$ from the finite structure the cuts are replaced by poles due to the Fabry-P\'erot interference. For thick enough structures the poles of $r(\omega)$ related to the edge states are close to those of $r_{\infty}(\omega)$ and, therefore, can be easily distinguished.

We characterize each structure layer by a $2\times 2$ transfer matrix linking the amplitudes of the right- and left-going waves (denoted by $+$ and $-$, respectively) at the right layer edge  with those at the left one. For a single layer $j =\mathcal A$, $\mathcal B$ this matrix reads
\begin{equation}\label{eq:T1}
\hat{T}^{(j)}(\omega) = \frac{1}{t_j(\omega)}
\begin{bmatrix}
t_j^2(\omega) - r_j^2(\omega) \; & r_j(\omega) \\ - r_j(\omega) & 1
\end{bmatrix}
\:,
\end{equation}
where the single layer reflection and transmission coefficients $r_j$ and $t_j$ are given by Eq.~\eqref{rt} for a resonant layer $\mathcal A$ while, for a spacing layer $\mathcal B$ of the width $L$, they are $r_{\mathcal B}=0$ and $t_{\mathcal B}=\e^{\rmi q L }$.  The total transfer matrix of the structure $\hat{T}^{(N)}(\omega)$ is given by the product of the individual transfer matrices through $N$ periods. The reflection coefficient from the left reads~\cite{Ivchenko2005}
\begin{equation} \label{rTT}
 r_N(\omega) = - T^{(N)}_{-+}(\omega) / T^{(N)}_{--}(\omega)\:.
\end{equation}
 As follows from Eqs.~\eqref{rt} and~\eqref{eq:T1} the transfer matrix elements for a single layer have no poles except for the trivial one at $\omega = \omega_0 - {\rm i} \Gamma$. Hence, the pole $\Omega$ of the reflection coefficient can be found from the condition $T^{(N)}_{--}(\Omega) = 0$.
This condition allows the existence of light waves going away from the system in the absence of incident waves, and thus it indeed determines the eigenmodes. For real $\omega$ the reflectance and transmittance are bounded by unity. Therefore, all the pole frequencies $\Omega$  should have non-zero imaginary parts and the corresponding eigenstates decay in time.

The similar consideration can be applied for a semi-infinite structure. Its reflection coefficient is expressed in terms of the transfer matrix through one period $T^{(1)}$ as
\begin{equation}\label{rinf}
r_{\infty}(\omega) = \frac{\e^{\i k(\omega)D}-T^{(1)}_{++}(\omega)}{T^{(1)}_{+-}(\omega)}\:, \,
\end{equation}
where $\e^{\i k(\omega)D}$ is an eigenvalue of the matrix $\hat{T}^{(1)}(\omega)$ and the polariton wave vector $k(\omega)$ is  chosen to have positive $\Im k(\omega)$. 

 The poles of  the reflection coefficient~(\ref{rinf}) are found from
\begin{equation}\label{polecond}
{T^{(1)}_{+-}(\Omega) = 0\:, \quad |T^{(1)}_{--}(\Omega)| < 1} \:.
\end{equation}
 The first condition means that only the outgoing wave is present on the left side of the structure, while the second condition ensures that the eigenstate spatially decays inside the structure. Hence, the conditions~(\ref{polecond}) select modes attached to the left edge.
In order to find the right-edge modes one should replace the first condition in Eq.~\eqref{polecond} with $T^{(1)}_{-+}(\Omega) = 0$.

{\it Results and discussion.} First we briefly consider a structure with $b=1/2$ and two resonant layers in the period. Its unit cell can be chosen 
to have a center of symmetry. In this case $T^{(1)}_{+-} = -T^{(1)}_{-+}$, see e.g.  Ref.~[\onlinecite{Podd_Ivch}]; at the frequency $\omega$ of the possible pole of $r_\infty$, the off-diagonal
 elements of matrices $\hat{T}^{(1)}$ and
$\hat{T}^{(N)} = \hat{T}^{(1)}{}^N$
 are zeros, and according to Eq.~(\ref{rTT}) the reflection coefficient vanishes rather than has a pole and, thus, the edge states are absent. 
Concomitantly, in the structure with $b=1/2$ the eigensolutions $P_l(k,\varkappa)$ of the Hamiltonian~\eqref{eq:system2} can be chosen in such a way that $P_l(k,\varkappa)=P_l(k,-\varkappa)$. As a result, the integrand in Eq.~\eqref{eq:chn} is odd in $\varkappa$ and all the Chern numbers are zero.
The absence of  {\it radiative} edge states is characteristic for centrosymmetric optical lattices. The  conventional electronic lattices may have edge (zero-energy) modes even for a 
centrosymmetric unit cell, e.g. in the Su-Schrieffer-Heeger model with two sites in the unit cell~\cite{Sheng2013}.

 \begin{figure*}[t]
\includegraphics[width=0.99\textwidth]{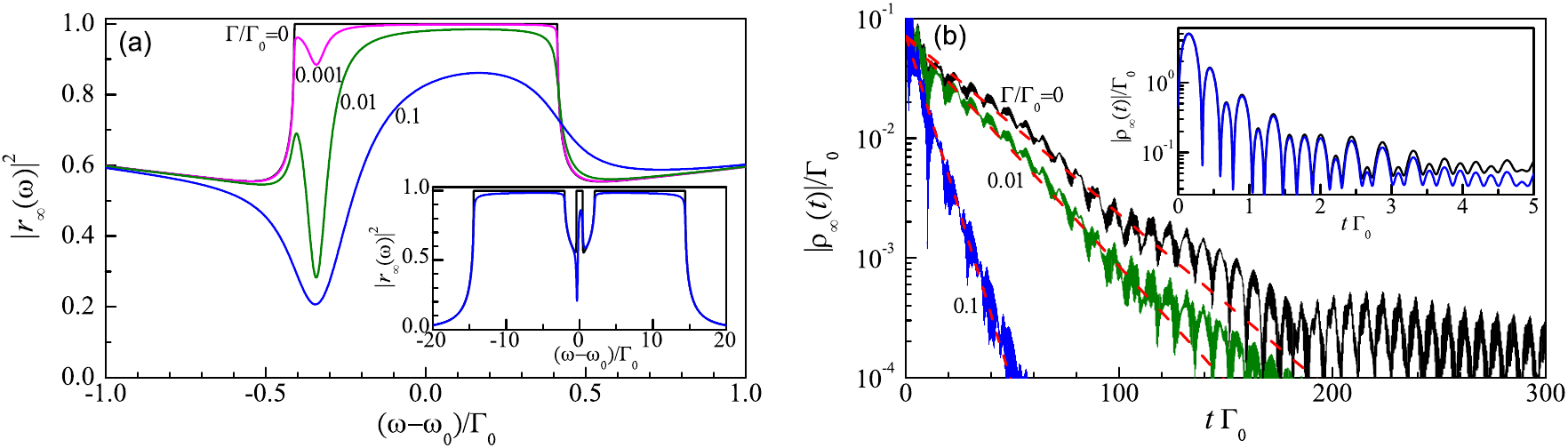}
\caption{(Color online) (a) The stationary reflection spectra $|r_\infty(\omega)|^2$ for the semi-infinite structure with $\varkappa=\phi-\pi/6=\pi/2$,  $\eta=0.2/\pi$, $\Gamma_0/\omega_0 =3\times 10^{-3}$ and various values of the non-radiative damping $\Gamma$. Inset presents the spectra in the wider spectral region showing the Bragg zones. (b) The short pulse response function $\rho(t)$ of the structure. Inset zooms the response function at the short time delays showing the contribution of Bragg zones. The dashed lines describe the edge mode contribution to the reflection and are plotted after Eq.~\eqref{rho}.
}
\label{fig:refl}
\end{figure*}

Now we turn to the lattice with $b=1/3$ comprising three resonant layers per period.  The dependence of the spectrum on the auxiliary wave vector component $\varkappa=\phi-\pi/6$ is presented in Fig.~\ref{fig:zones}.
We fix the attention on the narrow spectral range around the resonance frequency $\omega_0$ where the system has three allowed zones separated  by two band gaps (white areas)~\cite{Voronov2004}. The bands are $2\pi/3$-periodic in agreement with the discussed symmetry property of the Bloch states. Each of the three indicated Chern numbers differs from 0 and their sum gives zero. As a consequence the structure possesses two edge modes with the energies in the band gaps. The real parts of the mode eigenfrequencies are shown by the lines in Fig.~\ref{fig:zones}(a): the solid curve corresponds to the  mode localized on the left edge,  while the dashed curve corresponds to the right-edge mode. Figure ~\ref{fig:zones}(a) demonstrates that the edge modes traverse the band gaps when the parameter $\varkappa$ is varied from $-\pi$ to $\pi$. Note that, in contrast to the free Bloch solutions consistent with the translational symmetry for the index $n$, the $2\pi/3$-periodicity does not hold for the edge modes. However the inversion symmetry $\varkappa \to -\varkappa$ is retained and, indeed, swaps the left- and right-edge modes. The fact that the edge states for the values $\varkappa$ and $- \varkappa$ are localized at opposite interfaces reflects the ``topological protection'' of the lattice $(n,m)$.

Since the optical lattice is open the edge eigenmodes are non-stationary. The imaginary part of eigenfrequencies and the edge-mode spatial decay per unit cell, $\Im \{ k(\Omega)\} D$, are shown in Fig.~\ref{fig:zones}(b) and Fig.~\ref{fig:zones}(c), respectively. Figure~\ref{fig:zones} demonstrates that the structure has two edge eigenmodes for all values of $\varkappa$ excepting six special points. Particularly, for $\varkappa=0$~and~$\pi$ both edge states vanish: $\Im \{ k(\Omega)\} D \to 0$, $\Im \Omega \to 0$. This can be understood taking into account that, for these particular values of $\varkappa$, the structure is invariant under the reversal $\varkappa \to -\varkappa$ and hence centrosymmetric. For the other four special points $\varkappa=\pm\pi/6$~and~$\pm 7\pi/6$, the spacing between two adjacent resonant layers $\mathcal A$ equals to $\lambda_0/2$ and, as a consequence, one of the edge states disappears. Interestingly, the real part of $\Omega - \omega_0$ (as well as ${\rm Re}\{ k(\Omega)\}$) reverses the sign and the imaginary parts of $\Omega$ and $k(\Omega)$ are invariant under the shift of $\varkappa$ by $\pi$. This is a special property of the structure with $d = \lambda_0/2$ which can be proved by the complex conjugation of Eq.~(\ref{qw}).

For the most values of $\varkappa$ the edge modes are well-defined and localized within a few structure periods. They are already distinguishable in thin structures, as soon as the condition $|\exp{[\i N k(\Omega)D ]}| \ll 1$ is fulfilled. Fig.~\ref{fig:states} shows the spatial distribution of the absolute value of the edge-mode electric field for the five-period structure with $\varkappa=\pi/2$. Since the edge states are never degenerate, they  do not intermix even in finite structures. 

Now that we have demonstrated the existence of edge states we focus on the problem of their detection. For example, we consider the semi-infinite structure with $\varkappa=\pi/2$ marked by the star in Fig.~\ref{fig:zones}. For this value of $\varkappa$ the central allowed band shrinks and the values of ${\rm Re}(\Omega - \omega_0)$ have opposite signs for the left- and right-edge modes, see Fig.~\ref{fig:zones}. Under the assumption 
$\Gamma_0\ll\omega_0$ the left-edge mode frequency reads
 \begin{equation}\label{eq:edgefreq}
  \Omega =  \omega_0  - \frac{2\Gamma_0 \sin (3\pi \eta/2)}{\sqrt{2+\e^{-3\i \pi \eta}}} - \i\Gamma \:.
 \end{equation} 
The reflection spectrum  $|r_{\infty}(\omega)|^2$ for the semi-infinite structure is shown in Fig.~\ref{fig:refl}(a). The black curve corresponds to the absence of the non-radiative damping, $\Gamma=0$. In this case the edge state does not reveal itself in the spectrum. This is because its frequency is located in the band gap where the reflectivity is already equal to unity, and only the phase of the amplitude coefficient $r_{\infty}(\omega)$ is sensitive to the edge mode. When $\Gamma>0$ the edge state shows up as a dip in the reflection spectrum and a peak in the absorption spectrum.
Similar  approach has been used in Refs.~[\onlinecite{dyakov2012}, \onlinecite{tikhodeev2013b}] to detect conventional Tamm states~\cite{vinogradov2010,kaliteevski2007} in 2D centrosymmetric photonic crystals. The position and halfwidth of the reflectivity dip are determined, respectively, by the real and imaginary parts of $\Omega$. The inset in Fig.~\ref{fig:refl}(a) shows the reflectivity in a wider range of frequencies. One can see two Bragg stop-bands with borders at the frequencies $\sqrt{\omega_0 \Gamma_0 (1\pm |f|) /\pi}$, where $f$ is the structure factor~\cite{Podd_Ivch} which reduces, for small values of $\eta$, to $|f| = 1 - (\pi \eta)^2$. The detuning of these zones from $\omega_0$ exceeds by far the value of $\Gamma_0$, and they lie outside the spectral range of the edge states shown in Fig.~\ref{fig:zones}.

An alternative method of detecting the edge modes is the time-domain optical spectroscopy. The system can be described by the time-resolved reflection response  $\rho(t) = \int_{-\infty}^{\infty} r(\omega) \exp (-\i\omega t) d\omega/(2\pi)$ induced by  the short $\delta$-pulse~\cite{Poshakinskiy2012}. Such technique is sensitive both to the amplitude and phase of the reflection coefficient $r(\omega)$. The edge state should reveal itself as an exponential contribution to the response function given by the residue of $r_\infty (\omega)$ at the frequency $\Omega$,
\begin{equation}\label{rho}
\rho_\Omega (t) =  - \frac{ (1-\e^{3 \i \pi \eta })^2}{(1+2 \e^{3 \i \pi \eta})^{3/2}}\,\Gamma_0\,\e^{-\i \Omega t} \:.
\end{equation} 
Thus, the information about the phase missing in $|r(\omega)|^2$ shows up in $\rho(t)$. In Fig.~\ref{fig:refl}(b) the response function $\rho_\infty (t)$ is presented in the semi-logarithmic scale. It indeed contains an exponentially decaying contribution that perfectly agrees with Eq.~\eqref{rho}, see dashed curves. This contribution is already present for $\Gamma = 0$ although the edge state is not revealed in the  stationary spectrum $|r(\omega)|^2$, cf. the corresponding curves in Figs.~\ref{fig:refl}(a) and~\ref{fig:refl}(b). At longer times the exponential decay due to the edge state is masked by the $t^{-3/2}$ power-law contribution of the Bloch-states continuum, the black curve in Fig.~\ref{fig:refl}(b)~\cite{Poshakinskiy2012}.

To summarize, we have demonstrated the presence of radiative topologically protected edge states in the 1D resonant photonic crystals with a compound unit cell.  The edge states are shown to survive despite the long-range light-induced coupling of the resonances and the finite lifetime of their radiative decay.  Analytical transfer-matrix conditions for existence of edge states have been formulated. The appearance of the right- and left-edge states has been interpreted in terms of the topological properties of the  two-dimensional ``ancestor'' lattice obtained by the extension of the considered 1D lattice into the dual space. The states are manifested in the stationary reflection spectra of the structure with finite nonradiative losses as well as in  the time-dependent response to the short optical pulse. The plasmonic lattices characterized by  high enough radiative decay rate $\Gamma_0$ are preferential for the observation of edge states~\cite{taubert2012}.

\paragraph*{Acknowledgments.} 
The authors acknowledge fruitful discussions with S.~A.~Tarasenko.
This work was supported by the RFBR,  
RF President Grants MD-2062.2012.2 and NSh-5442.2012.2, EU projects ``SPANGL4Q'' and
``POLAPHEN'', and the Foundation ``Dynasty''.

\end{document}